\documentclass[10pt,twocolumn,twoside]{IEEEtran}
\newcommand{\comm}[1]{}

\def\noarXiv{\comm}
\usepackage{amssymb,epsfig}
\usepackage{amsthm}
\usepackage[numbers]{natbib}
\def\citet{\cite}
\usepackage{etoolbox}
 \usepackage{color}

\newtheorem{theorem}{Theorem}
\newtheorem{lemma}{Lemma}
\newtheorem{proposition}{Proposition}
\newtheorem{condition}{Condition}
\newtheorem{remark}{Remark}
\newtheorem{example}{Example}

\def\e{\varepsilon}

\def\defi{\stackrel{{\scriptscriptstyle \Delta}}{=}}

\def\o{\omega}
\def\O{\Omega}

\def\Y{{\cal Y}}

\def\w{\widehat}
\def\Ind{{\,\rm Ind\,}}
\def\Ind{\mathbb{I}}

\def\R{{\bf R}}

\def\Z{{\cal Z}}

\def\C{{\bf C}}

\def\ww{\widetilde}
\def\X{{\cal X}}

\def\oo{\bar}

\def\D{{\Delta}}

\def\V{{\cal V}}

\def\M{{\cal M}}

\newcommand{\be}{\begin{equation}}
\newcommand{\ee}{\end{equation}}
\newcommand{\bd}{\begin{displaymath}}
\newcommand{\ed}{\end{displaymath}}
\newcommand{\ba}{\begin{array}{ll}}
\newcommand{\ea}{\end{array}}
\newcommand{\baa}{\begin{eqnarray}}
\newcommand{\eaa}{\end{eqnarray}}
\newcommand{\baaa}{\begin{eqnarray*}}
\newcommand{\eaaa}{\end{eqnarray*}}
\font\sm=cmr10

\def\oo{\bar}

\def\BL{{\scriptscriptstyle BL}}

\def\sinc{{\rm sinc\,}}
\def\ew{\left(e^{i\o}\right)}
\def\T{{\mathbb{T}}}
\def\ZZ{{\mathbb{Z}}}

\def\TT{{\D}}
\def\XN{\ell_2^\BL}
\def\XNL{\ell_2^\BL(-\infty,0)}
\def\XNL{\ell_2^\BL(D)}
\def\BN{{\cal B}}
\def\BNN{{\cal B}}
\def\BN{{\mathbb{B}}}

\def\M{M}
\def\D{D}
\def\BN{{L_2^\BL(\T)}}
\def\BNN{{L_2^\BL(\T)}}
\date{Submitted April 29, 2\X^M016. Revised October 28, 2017}
\title{Optimal data recovery and forecasting with dummy long-horizon forecasts
}
\author{
Nikolai Dokuchaev }
\begin{document}
\def\break{}%
\def\brea{}
\def\breakk{}
\def\brea{\nonumber\\ }\def\breakk{\nonumber\\&&} 
\vspace{1cm}
\maketitle
\let\thefootnote\relax\footnote{Submitted April 29, 2016. Revised October 27, 2017.} 
\let\thefootnote\relax\footnote{ The author is with
 { Department of Mathematics \& Statistics, Curtin
University,} {  GPO Box U1987, Perth, 6845 Western Australia.} }
 \begin{abstract} The paper suggests  a method of recovering missing values
 for sequences, including sequences with a multidimensional index,
 based on optimal approximation by  processes featuring spectrum degeneracy.
The problem is considered in the pathwise setting, without using  probabilistic assumptions on the ensemble. The method requires to solve a closed linear equation  connecting
the available observations of the underlying process with the  values of the approximating
process with degenerate spectrum  outside the observation range.
 Some robustness with respect to noise contamination is established for the suggested recovering algorithm.
It is suggested to apply this data recovery algorithm to forecasting with  a preselected dummy long-horizon forecast that  helps to regularize the solution.
\end{abstract}

\vspace{-0.5cm}
\section{Introduction} The paper considers data recovery problems for sequences in pathwise setting, i.e.
without probabilistic assumptions, using the approach suggested in  \cite{D17a,D17b} in an extended setting,
allowing  more general type  of approximating processes, more general domains for missing data, and more general
sequences with a multidimensional index.

For continuous data, the recoverability is associated with smoothness or analytical properties of
the processes. For
discrete time processes,  it is less obvious how to interpret analyticity; so far, these problems
were studied in a stochastic setting,
where an observed process is deemed to be  representative of an ensemble of paths with
the probability distribution that is either known or can be estimated from repeating experiments.
For stochastic  stationary  Gaussian processes
with the  spectral density $\phi$, a classical result is that
a missing  single value is recoverable  with zero
error if and only if \baa
\int_{-\pi}^\pi \phi\ew^{-1} d\o=-\infty.\label{Km} \eaa
(Kolmogorov \cite{K},
Theorem 24). Stochastic stationary Gaussian processes without this property are called {\em minimal } \cite{K}. In particular, a process is recoverable
if it is ``band-limited" meaning that the spectral density is vanishing on an arc of the
unit circle $\{z\in\C:\ |z|=1\}$.
This illustrates how recoverability  is connected with  bandlimitiness or its relaxed versions.
Respectively, it is common to use band-limited approximations of non-bandlimited
underlying processes for the forecasting and other applications.
 There are many works devoted to causal smoothing and sampling, oriented on estimation and minimization of norm of the error, especially in stochastic setting; see e.g.
\citet{Alem,Cai,CTao,CJR,D12a,D12b,D16a,D16b,Donoho,F94,PFG,PFG1,FKR,jerri,K,LeeF,Peller,Pou,Pou2,rema,Sz,Sz1,T,TH,W,Zhao, Zhao2}.
  \index{\citet{jerri}, \citet{PFG}, \citet{PFG1}, \citet{AU}.} 
Some analogs of criterion (\ref{Km}) or error-free recoverability were obtained in \citet{D16b}.

The present paper considers optimal  data recovering problem  for  sequences that are not necessary
parts of band-limited processes. We consider the problem  in the deterministic setting, i.e. pathwise.
This means that the method has to rely on the intrinsic properties of a sole underlying sequence without appealing
to statistical properties of an ensemble  of sequences.   An estimate of the missing  value
   has to be done based  on the intrinsic properties of  this sole  sequence and the observed values.
The paper suggests a method of optimal recovering  missing  values of sequences (discrete time processes) based on extension of the approach from
\citet{D17a,D17b}. The  optimality criterion is pathwise; it does  not involve an expectation  on a probability space.
In \cite{D17a}
band-limited extensions of one-sided sequences with one-dimensional index were considered. In \cite{D17b}, data recovery based on band-limited
approximations  was considered for sequences with one-dimensional index with finite number of missing values at consequent points.
In the present paper, we consider more general type  of approximating processes, more more general domains for missing data, and 
sequences with a multidimensional index that can be used in spatial processes and image analysis.

\noarXiv{We consider an approximation in $\ell_2$-norm rather than  matching  the values at
selected points.
This is different from  a setting in \citet{CJR,F94,FKR,LeeF}.
Our setting is closer to the setting from  \citet{PFG,PFG1,TH,Zhao}.
In \citet{PFG,PFG1},  a band-limited process representing
smoothed
underlying  continuous time process was used. In \citet{TH}, the problem  of minimization of the total energy  of the approximating bandlimited process
 within a given distance from  the original process smoothed by an ideal low-pass filter was considered.
 In \citet{F94},  an interpolation problem for absent sampling points  was considered in a setting with vanishing error. In \citet{Zhao},  extrapolation of a band-limited matching a finite number of points process was considered using special Slepian's type basis in the frequency domain.
We consider extrapolation in time domain, with  the main focus  on the minimization of an essential  non-vanishing error.
}

The  method requires to solve a convenient closed linear equation  connecting directly
the set of  observations of the underlying process with the set of recovered values of the approximating process
(equation (\ref{yAa}) in Theorem \ref{ThM} and equation (\ref{yAaR}) in Theorem \ref{ThMR} below).
The equations are finite dimensional for the case of a finite number of missing values.
Since the selection of the basis in the frequency domain
is not required, this allows to avoid  calculation of Slepian's type basis \citet{SP}.
 We established  solvability and uniqueness of the solution of the recovering problem.
Furthermore,  we established  numerical stability and robustness of the method  with respect to the input errors and data
 truncation. For the case of a large set of missing values, this would require to impose a
 penalty on the norm of the approximation process, i.e. to run a  Tikhonov regularization. We found that this regularization can be achieved
   with  an arbitrarily small modification of the optimization problem   (Theorem \ref{ThMR}).
 For the case of a small finite set of missing values, this regularization is not required.

We considered  approximations by band-limited processes (Section \ref{SecBL}), in the setting that is close to the setting
from \citet{D17a,D17b} but with more general type of the domain for the missing data.
In this case,  the recovery  equation (\ref{yAa}) is in the time domain. 

As an example of applications, we consider forecasting of sequences on a short horizon.  Approximation of underlying processes by smooth predicable processes
is a traditional forecasting tool.  In the present paper, we suggest to supplement regularization  by a
penalty on the growth of the solution by regularization via  including some  dummy long-horizon forecast and considering forecasting as
a data recovery problem, with missing data between current time and times  covered by this
long-term forecast. In other words, we have to replace the extrapolation of the past path by interpolation between the past path and this dummy long-horizon forecast. It appears that this helps to stabilize the numerical
solution similarly to the penalty on the growth. Of course, the choice of this  dummy long-term forecast
 have am impact on the short-horizon  forecast; however, we found that this impact is mild with appropriate choices of the horizon.
 The  sustainability
   of the  method is illustrated with  some  numerical experiments.

\section{Some definitions and background}

Let $\ZZ$  be the set of all integers.

Let $V$ be a Hilbert space, and let $n\in\ZZ$ be given, $n\ge 1$.

Let  $\D\subset \ZZ^n$ be a given  set, and let $\M=\ZZ^n \setminus \D$.

 \par
 For a set $G\subset \ZZ^n$,
we denote by $\ell_2(G,V)$ a Hilbert
space of  sequences $\{x(t)\}_{t\in G}\subset V$
such that
$\|x\|_{\ell_2(G,V)}=\left(\sum_{t\in G}\|x(t)\|_V^2\right)^{1/2}<+\infty$.

\def\XN{\Y}
\def\XNL{\Y(D)}\def\BL{{\scriptscriptstyle \Y}}

We will denote $\ell_2(G)=\ell_2(G,\R)$ and $\ell_2=\ell_2(\ZZ)$.

Let a mapping $\nu:\ell_2(\ZZ^n,V)\to \ell_2(\ZZ^n,V)$ be defined such that
 $\nu(x)(t)=x(t)$ for $t\in \D$ and   $\nu(x)(t)=0$ for $t\in \M$, i.e. $\nu(x)(t)=\Ind_{\{t\in \D\}}x(t)$.

Let $\X$ be a closed linear subspace of $\ell_2(\ZZ^n,V)$ such that
if $x\in \X$ then $\nu(x)\in\X$.

Let $\X(D)$ be the subset of $\X$ formed by the traces
$\{\w x(t)\}_{t\in D}$ for all sequences  $\w x\in\X$.

Let $\X^M$ be the linear subspace in $\X$ consisting of all
$x\in\X$ such that $x(t)=0$ for $t\notin M$.

We will consider $\X$, $\Y$,  and $\X^M$  as  Hilbert spaces provided with the norm
from $\ell_2(\ZZ^n,V)$. Similarly, we will consider  $\X(D)$ as a Hilbert space provided with the norm
from $\ell_2(D,V)$.

Let $\XN$ be a linear subspace in  $\X$, and let $\XNL$ be the subspace of $\X$ formed by the traces
$\{\w x(t)\}_{t\in D}$ for all sequences  $\w x\in\XN$.

We assume that $D$ and $\XN$ are selected such that the following condition holds.
\begin{condition}\label{condD}
\begin{enumerate}
\item
 For any $x\in\XNL$, there exists a unique $\w x\in\XN$ such that $\w x(t)=x(t)$ for $t\in D$.
 \item  $\XN$ is a closed subspace of $\ell_2(\ZZ^n,V)$.
\end{enumerate}
\end{condition}
Condition \ref{condD} implies that the trace $\{\w x(t)\}_{t\in M}$ of a
 process  $\w x\in\XN$
 is uniquely defined by its trace
$\{\w x(t)\}_{t\in D}$.

The setting of this paper targets situations where this condition is satisfied for classes $\Y$
of processes with some spectrum degeneracy, as a  generalization of the setting from \cite{D17a,D17b}, where
the case of $n=1$, $V=\R$, $\X=\ell_2$ was consider, with $\Y$ being a set of band-limited processes, in the notations of the present paper. 
In \cite{D17a}, finite sets $M$ were considered. In \cite{D17a}, the set  $D=\{t\in \ZZ:\ t\le 0\}$ was considered.

More general setting studied in the present paper  allows to cover a variety of models. For example, 
inclusion of $n>1$ allows applications for the image analysis and spatial processes.
Models with finite sequences can be covered  with the selection of $\X=\{x\in \ell_2(\ZZ^n,v):\ \|x(t)\|_V=0\quad \hbox{if}\quad |t|<N\}$, for $N>0$.

\section{The equation for optimal recovering}
We consider below input processes $x\in\X$ and their approximations in $\XN$.
The sequences $\{x(t)\}_{t\in \D}$ represent the available data;
the values for $t\in \M$ are unavailable.

We will
be using approximation described in the following lemma.
\begin{lemma}\label{lemma1}
\begin{enumerate}
\item $\Y(D)$ is a closed linear subspace of $\X$.
\item
There exists a unique optimal solution  $\w x\in\XN$
of the minimization problem \baa &&\hbox{Minimize}\quad  \sum_{t\in \D }\|x_\BL(t)-x(t)\|_V^2  \quad\breakk\hbox{over}\quad x_\BL\in \XN .\label{min} \eaa
\end{enumerate}
\end{lemma}

\par
Under the assumptions of Lemma \ref{lemma1},
there exists a unique process $\w x\in\XN$ such that the trace $\w x|_{t\in \D}$
provides an optimal approximation of
its observable trace  $\{ x(t)\}_{t\in \D}$.  The corresponding  trace  $\{\w x(t)\}_{t\in M}$  can
be interpreted  as a result of optimal recovery of missed trace $x|_M$ (optimal in the sense of problem (\ref{min}) given $\Y$). We will suggest below a method of finding
this trace $\{\w x(t)\}_{t\in M}$ only; the calculation of the trace $\{\w x(t)\}_{t\in \D}$ will not be required and will be excluded.

Let $P:\X\to\Y$ be the projection operator.

Let $A:\X^M\to \X^M$ be an operator defined as
\baa
Ay=\Ind_{M}P (\Ind_{M}y). \eaa

 Let  a mapping $a:\X(\D)\to \X^M$  be defined   as
\baa
a(x)=\Ind_{\M}P(\nu(x)). \eaa

\vspace{0.1cm}
\begin{theorem}
\label{ThM} For any $x\in\X(\D)$, the equation \baa y=Ay+a(x)
\label{yAa}
\eaa has a unique solution  $\w y=\Ind_{M} \w x\in \X^M$.
\end{theorem}
The trace $\w y|_{t\in\M}$ of the solution in Theorem \ref{ThM} is the sought extension  on $\M$ of the
optimal $\w x$ approximating the observed sequence $\{x(t)\}_{t\in \D}$.

Approach of Theorem \ref{ThM} represents a version of so-called alternative projection
method; see e.g. Theorem 4 in \cite{Donoho}, p. 912-913, and the 
bibliography therein.  It can be noted that  
the assumptions of Theorem \ref{ThM} do not require restriction on data sparsity or the wide
of the frequency  band, as is usually required by data recovery algorithms exploring 
uncertainty principle \cite{Donoho}.

 The following lemma shows  that the mapping $A$ is not a contraction but it is close to a contraction.
\begin{lemma}
\label{lemma3}
\begin{enumerate}
\item For any $y\in \X^M$ such that $y\neq 0$,  $\|Ay\|_{\X^M}<\|y\|_{\X^M}$.
\item The operator $A:\X^M\to \X^M$  has the norm $\|A\|\le 1$.
\item
If the space $V$ is finite dimensional and  the set $M$ is finite, then  the operator $A:\X^M\to \X^M$  has the norm $\|A\|<1$.
\end{enumerate}
\end{lemma}

\subsection*{Regularized setting}
Let us consider a modification  the original problem (\ref{min})
\baa &&\hbox{Minimize}\quad  \sum_{t\in \D}\|x_\BL(t)-x(t)\|_V^2 +\rho\|x_\BL\|_{\ell_2}^2 \quad\breakk\hbox{over}\quad x\in \XN. \label{minR} \eaa
Here  $\rho\ge 0$ is a parameter.

The setting  with $\rho>0$ helps to prevent selection of $\w x$ with excessive norm.
It can noted that it is common to
  put restrictions on the norm of the optimal process
  in data recovery, extrapolation, and interpolation problems in signal processing; see e.g. \citet{Alem,CTao,TH}.

Lemma \ref{lemma1} can be generalized as the following.
\begin{lemma}\label{lemma1R}
For any $\rho\ge 0$ and  $x\in \X(\D)$, there exists a unique optimal solution  $\w x_\rho$
of the minimization problem (\ref{minR}). In addition, $\w x_\rho=(1+\rho)^{-1}\w x_0$.
\end{lemma}
\par
In these notations, $\w x_0$ is the optimal process presented in Lemma \ref{lemma1}.\par

\par
Under the assumptions of Lemma \ref{lemma1R}, the
trace on $\M$ of the band-limited solution $\w x_\rho$ of problem (\ref{minR})  can
be interpreted  as a result of optimal recovery of the missed trace of $x|_{\M}$ (optimal in the sense of problem (\ref{minR})  given $\O$ and $\rho$).
Let us derive an equation for this solution.
\vspace{0.1cm}
\par

\par
Let $I:\X^M\to \X^M$ be the identity operator.

 Let
$A_\rho=(1+\rho)^{-1}A$ and $a_\rho(x)=(1+\rho)^{-1}a(x)$, where  $A$  and $a(x)$ are such as  defined  above.

\par
 It follows immediately from Lemma \ref{lemma3}(ii) that, for any $\rho>0$, $\|A_\rho\|<1$.
 Hence the operator $\left(I-A_\rho\right)^{-1}: \X^M\to \X^M$ is continuous
 and
 \baaa
 \left\|(I-A_\rho)^{-1}\right\|<+\infty,
  \eaaa
  for the corresponding norm. In addition, by the properties of projections presented in the definition  for $a(x)$, we have that $\|a_\rho(x)\|_{\X^M}\le \|x\|_{\X(D)}$.

Theorem \ref{ThM} stipulates  that equation (\ref{yAa}) has a unique solution.  However, this theorem does not establish
the continuity of the dependence of $\w y$ on the input $x|_{t\in \D}$.
The following theorem shows that regularity of solutions is feasible  for problem  (\ref{minR}) with $\rho>0$.
 \begin{theorem}
\label{ThMR}  \begin{enumerate}
\item
For any $\rho> 0$ and $x\in\X(D)$, the  equation \baa
(1+\rho)y=Ay+a(x)
\label{yAaR}
\eaa  has a unique solution   $y_\rho=\Ind_{\M}\w x_\rho\,=(I-A_\rho)^{-1}a_\rho(x)$ in $\X^M$. Furthermore,
for any $\rho>0$,
\baa
\|y_\rho\|_{\ell^M_2}\le \left\|(I-A_\rho)^{-1}\right\| \|x\|_{\X(D)}
\label{eest}
\eaa
for any $x\in\X(D)$.
\item If the set $M$ is finite and the space $V$ is finite dimensional, then statement (i) holds for $\rho=0$ as well.
\end{enumerate}
\end{theorem}
\par
Similarly to Theorem \ref{ThM}, the trace $\w y_\rho|_{t\in\M}$ of the solution in Theorem \ref{ThMR} is the sought extension  on $\M$ of the
optimal band-limited $\w x$ approximating the observed sequence $\{x(t)\}_{t\in \D}$ (optimal in the sense of problem (\ref{minR}) given $\O$ and $\rho$).

Replacement of the original problem by problem (\ref{minR}) with $\rho\to 0$ can be regarded as a Tikhonov regularization of the original problem. By Theorem \ref{ThMR}, it
 leads to solution  featuring continuous dependence on $x$ in the corresponding  $\ell_2$-norm.

\begin{remark}\label{remIT} Since the operator $A_\rho$ is a contraction, the solution of (\ref{yAaR}) can be approximated by  partial sums  $\sum_{k=0}^dA_\rho^ka_\rho(x)$.
\end{remark}

\section{Numerical stability and robustness}
 Let us consider a situation where an input process $x\in\X(D)$ is observed with an error.
 In other words, assume that we observe a process $x_\eta=x+\eta$, where $\eta  \in\X(D)$ is a noise.
 Let $y_\eta$ be the corresponding solution of equation (\ref{R})  with $x_\eta$ as an input, and let $y$
 be the corresponding solution of equation (\ref{R})  with $x$ as an input. By Theorem \ref{ThMR}, it follows immediately that, for $\e>0$,
  \baaa
  \|y-y_\eta\|_{\X^M}\le \left\|(I-A_\e)^{-1}\right\| \|\eta\|_{\X(D)}\quad \hbox{for all}\quad \eta\in \X(D).
  \eaaa
  This demonstrates some robustness of the method  with respect to  the noise in the observations.

In particular, this ensures robustness with respect to truncation of the input processes,
such that semi-infinite sequences $x\in \X(D)$ are replaced by truncated sequences $x_\eta(t)=x(t)\Ind_{\{|t|\le q\}}$ for $q>0$; in this case
   $\eta(t)=\Ind_{|t|> q}x(t)$ is such that  $\|\eta\|_{\X(D)}\to 0$ as $q\to +\infty$.
 This  overcomes principal impossibility to access infinite sequences of observations.

 In practice,  only  finite-dimensional systems of linear equations can be solved numerically.
 This means that, in the case where the set $M$ is infinite, equation (\ref{yAaR})   cannot be solved numerically even for truncated inputs, since
  it involves  a sequence $a(x)$ that has an infinite support for truncated  $x$.
   Therefore, we have to apply the method with $A$ replaced by its truncated version represented by
   a matrix of finite dimension.
   We will consider below the impact of  truncation of  $A$.
 \subsubsection*{Robustness with respect to the data errors and truncation}
For $N\in\ZZ$, $N>0$, let $D_N=\{t:\ |t|\le N\}$, and let the operator $A_N:\X^M\to\X^M$ be defined as
\baaa
A_Ny=\Ind_{M\cap D_N}P (\Ind_{M\cap D_N}y). \eaaa
Replacement of $A$ by $A_N$  addresses the restrictions on the data size for numerical
methods.
\par
   Again, we consider a situation where an input process  is observed with an error.
 In other words, we assume that we observe a process $x_\eta=x+\eta \in\X(D)$, where $\eta \in\X(D)$ is a noise.
 As was mentioned above, this allows to take into account truncation of the inputs as well.
\begin{lemma}
\label{lemma4} For any $N>0$, the following holds.
\begin{enumerate}
\item $\|A_Ny\|_{\X^M}\le \|y\|_{\X^M}$ for any $y\in \X^M$.
 \item For any $\rho\ge 0$ and any $x\in\X(D)$, the equation \baa (1+\rho)y=A_Ny+a(x)
\label{eqN}
\eaa has a unique solution  $\w y\in \X^M$.
\end{enumerate}
\end{lemma}

  \begin{theorem}\label{Th3} For  $\rho>0$, the solution of (\ref{yAaR}) is robust with respect to data errors and truncation, in the sense that
    \baaa
    \|y_{\rho,\eta,N}-y_\rho\|_{\X^M}\to 0\quad \hbox{as}\quad N\to +\infty\ \ \hbox{and}\ \ \|\eta\|_{\X(D)}\to 0 .
    \eaaa
 Here  $y_\rho$ denote  the solution  in $\X^M$ of equation (\ref{yAaR}), and  $y_{\rho,\eta,N}$ denote the solution in $\X^M$ of equation (\ref{eqN})  with $x$ replaced by $x_\eta$,
 such that $x\in\X(D)$, $\eta\in\X(D)$, and  $x_\eta=x+\eta$.
 \end{theorem}
 Theorem \ref{Th3} establishes robustness with respect to truncation of $(A,x)$ and with respect to the presence of the noise in the input.
 Therefore, this theorem  justifies acceptance of a result for $(A_N,x_\eta)$ as an approximation of the sought  result for $(A,x)$.
\section{Special case: approximation by band-limited processes}
\label{SecBL}
\def\XN{\ell_2^\BL}
\def\XNL{\ell_2^\BL(D)}
\def\BL{{\scriptscriptstyle BL}}

In this section, we assume that $V=\R$, $n=1$, and $\X=\ell_2=\ell_2(\ZZ,\R)$.

For  $x\in \ell_2$, we denote by $X=\Z x$ the
Z-transform  \baaa X(z)=\sum_{t=-\infty}^{\infty}x(t)z^{-t},\quad
z\in\C. \eaaa Respectively, the inverse $x=\Z^{-1}X$ of Z-transform   is
defined as \baaa x(t)=\frac{1}{2\pi}\int_{-\pi}^\pi
X\left(e^{i\o}\right) e^{i\o t}d\o, \quad t=0,\pm 1,\pm 2,....\eaaa

We assume that we are given $\O\in(0,\pi)$.
\par
Let  $\T=\{z\in\C:\ |z|=1\}$.

Let $\BNN$ be the set of all mappings $X:\T\to\C$ such
that $X\ew \in L_2(-\pi,\pi)$ and $X\ew =0$ for $|\o|>\O$. We will call  the corresponding processes $x=\Z^{-1}X$
{\em band-limited}.

Let $\XN$ be the set of all band-limited processes from
$\ell_2=\ell_2(\ZZ,\R)$, and let $\XNL$ be the subset of $\ell_2(D)=\ell_2(D,\R)$ formed by the traces
$\{\w x(t)\}_{t\in D}$ for all sequences  $\w x\in\XN$.

We will use the notation $\sinc(x)=\sin(x)/x$, and we will  use notation ``$\circ$''  for the convolution in $\ell_2$.

Let $H(z)$ be the transfer function for an ideal low-pass filter such that $H\ew=\Ind_{[-\O,\O]}(\o)$, where
$\Ind$ denotes the indicator function. Let $h=\Z^{-1}H$;
it is known that  $h(t)=\O\,\sinc(\O t)/\pi$.
The  definitions imply that $P x\in \XN$ for any $x\in \ell_2$ and that $P x=h\circ x$.

\begin{proposition}\label{propU}  Let $\Y=\XN$, and let $\D\subset \ZZ$ be such that there exist $s\in\ZZ$
such that either $\{t:\ t\le s\}\subset \D$ or $\{t:\ t\ge s\}\subset \D$. Then Condition \ref{condD} holds.
\end{proposition}
\par
Proposition \ref{propU}  can be considered as  reformulation in the deterministic setting
of a sufficient condition of predictability  implied by the classical Szeg\"o-Kolmogorov Theorem known for stationary Gaussian processes
\citet{K,Sz,Sz1}.

Up to the end of this section, we assume that $\D$ and $\Y$ are  such as described in Proposition \ref{propU}.

By the definitions, we have that
\baa
Ay=\Ind_{\M}(h\circ y), \quad  a(x)=\Ind_{\M}\left(h\circ(\nu(x))\right). \eaa
\par
  Since $h(t)=\O\,\sinc(\O t)/\pi$, the operator  $A= \Ind_{\M}(P \cdot)$ can be represented  as a matrix with the components \baaa A_{t,m}
=\Ind_{\{t,m\in M\}}\frac{\O}{\pi}\sinc[\O (t-m)], \quad t,m\in\ZZ,
\label{A}\eaaa
and $a(x)=\{a(x,t)\}_{t\in\ZZ}$ can be represented  as a vector  \baaa a(x,t)=\Ind_{\{t\in\M\}}\frac{\O}{\pi}\sum_{m\in D} x_m \sinc[\O(t-m)],\quad t\in\ZZ.\label{a}\eaaa

\begin{example}
\label{Th1} {\rm Let $s\in\ZZ$ be given. If $M=\{s\}$  is a singleton, then the problem becomes the problem of optimal recovery of  a missing value $x(s)$
for $x\in\X(D)$, where $\D=\ZZ\backslash \{s\}$.
 By Theorem \ref{ThM},
the problem
has an unique solution given
 \baa
\w x(s)=
\frac{\O}{\pi-\O} \sum_{m\neq s} x(m) \sinc[\O (s-m)].
\label{wx1}
\eaa
(The same solution was obtained in \cite{D17b}). This solution is optimal in the sense of problem (\ref{min}) given $\O$, with $M=\{s\}$.
In addition,
\baaa
|\w x(0)|\le \frac{\O}{\pi-\O} \|x\|_{\X(D)}.
\eaaa
To obtain this, we have to apply Theorem \ref{ThM}  with $M=\{s\}$. We have that the mapping $y\to y(s)$  is  a bijection between $\X^M$ and $\R$, $A_{t,m}=\frac{\O}{\pi} \Ind_{\{t=s,m=s\}}$, and optimal solution of the recovering problem is
 $y=\{y(t)\}_{t\in\ZZ}$ is such that $y(t)= 0$ for $t\neq s$, and
 \baaa
 y(s)= \frac{\O}{\pi} y(s)+\frac{\O}{\pi}\sum_{m\neq 0} x(m) \sinc[\O(s-m)]
 \eaaa
 or
 \baa
y(s)=\left(1-\frac{\O}{\pi}\right)^{-1} \frac{\O}{\pi}\sum_{m\neq s} x(m) \sinc[\O(s-m)].
\eaa
 It gives equation (\ref{wx1}) for the result $\w x(s)=y(s)$ of the optimal recovering of the missing value $x(s)$. $\Box$
 }\end{example}
\begin{remark}\label{corr1} Formula (\ref{wx1}) applied to $x_\BL\in \XN$ gives that f
 \baaa
x_\BL(s)=
\frac{\O}{\pi-\O} \sum_{m\neq s} x_\BL(m) \sinc[\O (s-m)]
\label{wx2}
\eaaa
This formula is known \citet{F92,F94a}; however,  equation (\ref{wx1})
 is different  since  $x$ in (\ref{wx1}) is not necessarily band-limited.
\end{remark}
\begin{example}
\label{ex2} {\rm  If $M=\{0,1,2\}$, then the problem becomes the problem of optimal recovery of  a missing values $(x(0),x(1),x(2))$.
 In this case,
 the result $\{\w x(t)\}_{t=0,1,2}= \{y(t)\}_{t=0,1,2}$ of the optimal recovering is
 $(I-\oo A)^{-1}\oo a$, where $I$ is the unit matrix in $\R^{3\times 3}$,
 \baaa
 &&\oo A=\{A_{t,m}\}_{t,m=0}^3\breakk = \frac{\O}{\pi}   \left(
                            \begin{array}{ccc}
                              1 & \sinc(\O ) & \sinc(2\O)\\
                              \sinc(\O ) & 1 & \sinc(\O)\\
                              \sinc(2\O ) & \sinc(\O ) & 1 \\
                            \end{array}
                          \right)\in\R^{3\times 3},
 \\
 &&\oo a=\{a(x,t)\}_{t=0,1,2}\in\R^3,\qquad\breakk a(x,t)=\frac{\O}{\pi}\sum_{m\notin\{0,1,2\}} x_m \sinc[\O(t-m)].\label{ooa}\eaaa
It was shown in \cite{D17a} that the matrix $I-\oo A$ is invertible.}
\end{example}

\subsection{Example of an application: forecasting with  a dummy long-horizon forecast}
Assume that   $D=\ZZ^-=\{t\in\ZZ:\ t\le 0\}$, i.e. that we observe past values $\{x(t)\}_{\{t\le 0\}}$ and wish to predict the values at $t\in  \ww M$, where $\ww M=\{1,...,\ww m\}$
using the extension $\w x|_{\ZZ^+}$ of a band-limited approximation of $x|_{\ZZ^-}$, i.e., solution of
problem (\ref{min}) with $\D=\ZZ^-$ and $M=\ZZ^+$.  In this case, we have not established the continuity  of the operator $(I-A)^{-1}$. Moreover,
our numerical experiments show that the minimal eigenvalues of the truncated operator $I-A_N$ converges to zero as $N\to +\infty$.
To ensure numerical stability and exclude excessive growth of $\w x(t)$ for $t>0$, we may use a modified problem (\ref{minR}) with $\rho>0$.
Let $\w x_\rho$ be the corresponding solution. Unfortunately this approach always leads to decreasing of the norm $\w x_\rho|_{\ZZ^-}$, so the approximation error
$\|x- \w x_\rho\|_{\ell_2(\ZZ^-)}$ can be large.

We suggest  an alternative approach. We suggest the following algorithm for predicting of  $x|_M$:
\begin{enumerate}
\item Select $m>\ww m$. Set $M=\{1,...,m\}$ and $\D=\ZZ\backslash M=\ZZ^-\cup \ww \D$.
\item Select  some  dummy sequence $z\in\ell_2(\ww \D)$ as a dummy long-horizon forecast. Here $\ww D\defi \{t\in\ZZ: t> m\}$.
\item Find $\w x\in \XN$ solving problem (\ref{min}) or (\ref{minR}) with some $\rho>0$.
\item Accept $\w x(t)$ as an optimal short-horizon forecast for $t=1,...,\ww m$ .
\end{enumerate}

It can be noted that, for finite $M$ and finite dimensional $V$, the operator $(I-A)^{-1}:\X^M\to \X^M$ is continuous. Hence we can select $\rho=0$
in (\ref{minR}). This approach could be preferable since it does not penalize directly for a large norm of $x_\BL|_D$  regardless of the choice of $z$. On the other hand,
the choice of $\rho>0$ leads to a distortion of the performance criterion since it penalizes for a large norm of $x_\BL|_D$.  However,
our numerical experiments show that the minimal eigenvalues of operator $I-A$ is quite small, especially  if
the number of elements of $M$ is large. This is the reason why we may use solution of (\ref{minR}) with some small $\rho>0$, to achieve
more robust numerical stability.

In some numerical experiments described below, we found that  if $\ww m$ is significantly smaller than  $m$ then
the choice of a dummy long-horizon forecast $z|_{\ww\D}$ has a relatively weak impact on the short-horizon forecast  $\w x|_{\ww M}$.

This can be explained as the following.

Let $\w z\in\ell_2(\ZZ\setminus \ZZ^-)$ and $\ww m>0$ be given, and let $z\in \ell_2(\ww D)$ be selected such that $z(t)=\w z(t-m+1)$, $t>m$.
We have that
\baaa
&&\w x=(I-A_\rho)^{-1}a(x)=(I-A_\rho)^{-1}(\Ind_{\{t\le 0\}}a(x))+\w x_m,\eaaa
where \baaa\w x_m=(I-A_\rho)^{-1}(\Ind_{\{t>m\}}a_\rho(x))=\brea
 (I-A_\rho)^{-1}(\Ind_{\{t>m\}}(P (\nu_m(z))).
\eaaa
Here $\nu_m(z)$ is an element of $\ell_2$ such that $\nu_m(z)(t)=z(t)$ for $t\ge m$ and  $\nu_m(z)(t)=0$ for $t<m$.
Clearly, for any $\w z$, we have that $(\Ind_{\{t>m\}}(P (\nu_m(z)))\to 0$ weakly in $\ell_2$ as $m\to +\infty$. Hence  $\w x_m\to 0$ weakly in $\ell_2$ as $m\to +\infty$.
It follows that $x_m(t)\to 0$   as $m\to +\infty$ for $t=1,..,m_0$. In other words,
\baaa
\w x|_{\ww M}\to (I-A_\rho)^{-1}(\Ind_{\{t\le 0\}}a(x))|_{\ww M}\quad\hbox{as}\quad m\to +\infty;
\eaaa
the limit here does not depend on $z$.
This implies that the short-horizon forecast obtained by this method can be similarly meaningful  for  different choices  of the dummy long-horizon forecast $z$.
We observed this feature in some numerical experiments described below.

In these experiments, we calculated
the solution $\w x|_M$ of linear system (\ref{yAa}) for a given $x$ directly using build-in MATLAB operation for solution of
linear algebraic
systems.

We used truncated input sequences $\{x(t)\}_{t\in\{-q,...,0\}}$ and  matrices
 $\{A_{t,m}\}_{k,m\in\{1,...,N\}}$, for  $q,N\in\ZZ^+$. We selected $N>m$ and $q=N$.
 The sequences were generated using Monte-Carlo simulation.
 The experiments demonstrated a good numerical stability of the  method; the calculations  were completed in few seconds; the results were
 quite robust with respect to deviations of input processes and truncation.
\par
\begin{figure}[ht]
\centerline{\psfig{figure=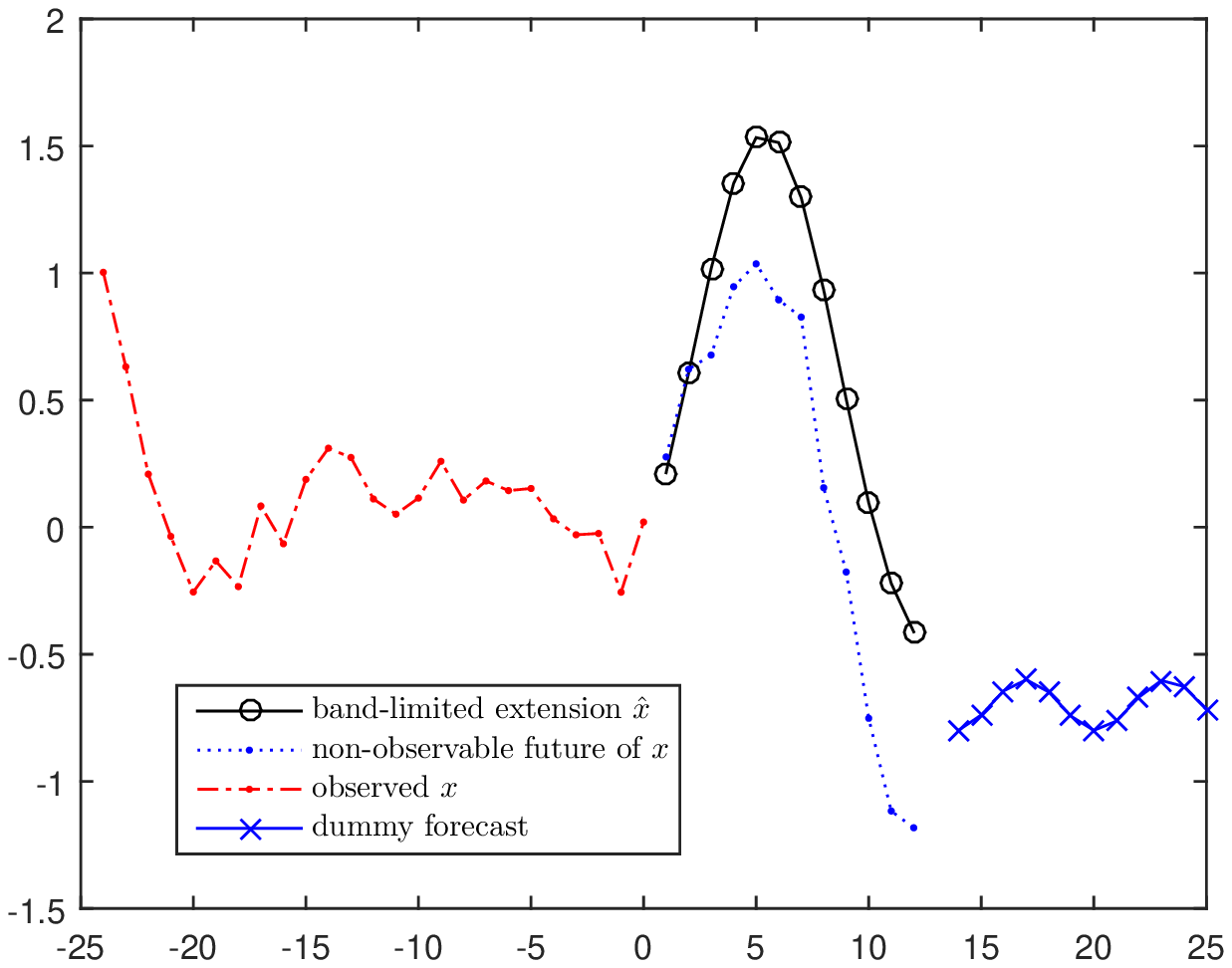,width=9cm,height=7.0cm}}
\centerline{\psfig{figure=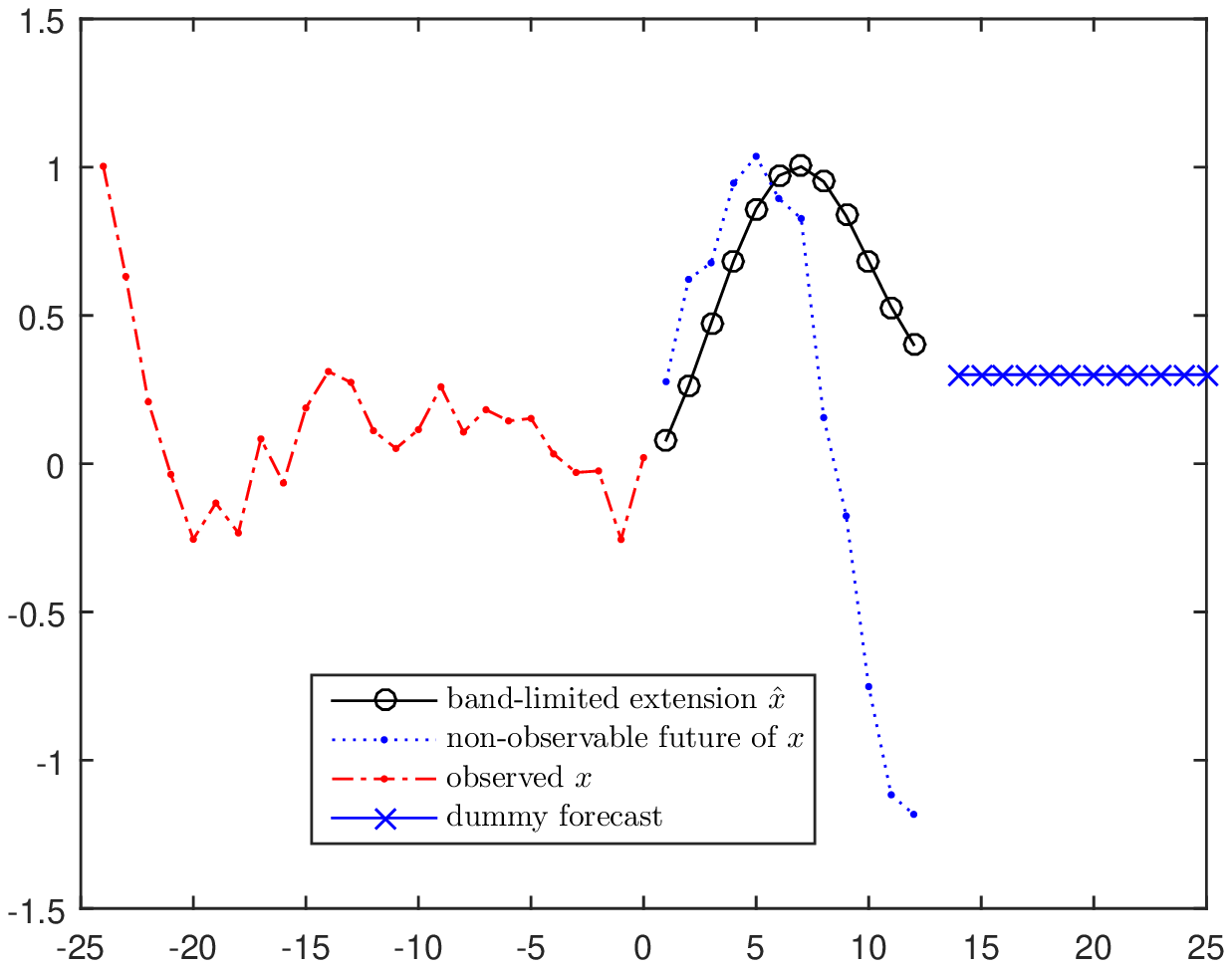,width=9cm,height=7.0cm}}
\caption[]{\sm
Examples of $x(t)$ and band-limited interpolations/forecasts    $\w x(t)$
for two different dummy long-term forecasts. } \vspace{0cm}\label{fig-1}\end{figure}

Figure \ref{fig-1} shows an  example of  a process $x(t)$, and examples
of the
corresponding band-limited extensions  $\w x|_{M}$ obtained  from (\ref{yAaR})
 with  $\O=0.25\pi$,  $q=-60$, $N=60$, $m=12$, $M=\{1,2,...,m\}$ with two different dummy sequences $z$ (i.e. dummy long-horizon forecasts).
Since  our method does not require
 $\w x(t)|_{t\notin M}$, these values were not calculated; the
extension $\w x(t)|_{t\in M}$ was derived directly from  $x(t)|_{t\le 0}$. Respectively, the values $\w x(t)|_{t\notin M}$
are not shown.
 \par
 It can be noted that the paths $\{\w x(t)\}_{1\le 1\le \ww m}$ are close if $\ww m$ is small.
\par
As was mentioned above, the extension $\w x|_{t>0}$ to the future times $t=1,..,\ww m$ can be interpreted as an
optimal forecast of $x|_{t\le 0}$ (optimal in the sense of problem (\ref{minR}) given $\O$ and $\rho$).

\section{Proofs}
 The following  proofs  represent extension of the  proofs given in \cite{D17a}
  for the case where $n=1$, $D=\ZZ^-$,  and $\V=\R$.
\def\XN{\Y}
\def\XNL{\Y(D)}
\def\BL{{\scriptscriptstyle \Y}}

 {\em
Proof  of Lemma \ref{lemma1}.} Consider the mapping $\zeta:\Y\to \Y(D)$ such that
$x(t)=(\zeta (x))(t)$ for $t\in D$. Clearly, it is a linear
continuous operator. By Condition \ref{condD}(i), it is a bijection.
Since  the mapping $\zeta:\XN \to \XNL$ is continuous, it follows that
the inverse mapping $\zeta^{-1}: \XNL\to\XN$ is also
continuous; see, e.g., Corollary in Ch.II.5 \citet{Yosida}, p. 77. Since the
set $\XN$ is a closed linear subspace of $\X$, it
follows that $\XNL$ is a closed linear subspace of $\X(D)$.
This completes the proof of statement (i). 

Further,  problem (\ref{min}) can be represented as 
problem \baaa &&\hbox{Minimize}\quad  \sum_{t\in \D }\|\oo x(t)-x(t)\|_V^2  \quad\breakk\hbox{over}\quad \oo x\in \XNL.\label{mina} \eaaa
By statement(i), there is an unique solution $\oo x\in \XNL$ of this problem; this is a projection of $x|_{D}$ on $\XNL$. 
Then a solution $\w x$  of problem (\ref{min})
is such that $\w x|_{D}$ is this $\oo x$, and this $\w x$ is unique by Condition \ref{condD}(ii).
 This completes the proof of  Lemma \ref{lemma1}.  $\Box$

\par
{\em Proof of Theorem \ref{ThM}.}  Let $\w x$ be the optimal solution described in Lemma \ref{lemma1}.
%
Let
$\X'=\{x\in\X:\  x|_{M}=\w x|_{D}\}$.
For any $x\in\X'$ and $\ww x_\BL\in \Y$, we have that
 \baaa
&& \|\w x-x\|_{\X}^2  =\|\w x-x\|_{\X(D)}^2+\|\w x-x\|_{\X^M}^2\\&&=\|\w x- x\|^2_{\X(D)}\le \|\ww x_\BL-x\|^2_{\X^M}.
\eaaa
The last inequality here holds because $\w x|_{D}$  is optimal for problem (\ref{min}).
This implies that, for any $x\in \X$, the sequence   $\w x$
is optimal for  the minimization problem \baaa &&\hbox{Minimize}\quad  \|
x_\BL-x\|_{\X}\quad\hbox{over}\quad x_\BL\in \XN.\quad\label{minPP} \eaaa
\par
Since $P:\X\to\XN$ is the projection operator, the optimal process $\w x\in \XN$ from Lemma \ref{lemma1} is such that
\baaa \w x=P x=P\left(\nu(x)+\Ind_{M}\w x \right). \eaaa
For $\w y=\Ind_{M}\w x$, we have that
\baaa \w y=\Ind_{M}\left( P\left(\nu(x)+\Ind_{M}\w x\right)\right) =
\Ind_{M} (P\nu(x)) +\Ind_{M}(P (\Ind_{M}\w x))\\=a(x) +A\w y. \label{xyM}\eaaa
This completes the proof  of Theorem \ref{ThM}. $\Box$
\par

\par
{\em Proof of Lemma \ref{lemma3}.} Let us prove statement (i). Let $y\in\X^M$ and $y\neq 0$. In this case, by Condition \ref{condD},
$y\notin \XN$.
Hence   $\|P y\|_{\X}< \|y\|_{\X}$.
Hence
\baaa
\|A y\|_{\X^M}=\|\Ind_{\M}P y\|_{\X}\le \|P y\|_{\X}< \|y\|_{\X}=\|y\|_{\X^M}.
\eaaa
This completes the proof of statement (i) of  Lemma \ref{lemma3}.
Statement (ii) follows from statement (i).
Statement (iii) follows from statement (i) and from  finite dimensionality of  $\X^M$ in this case.
This completes the proof of  Lemma \ref{lemma3}. $\Box$

\par
{\em Proof  of Lemma \ref{lemma1R}.}
As was shown in the proof of Lemma \ref{lemma1},  $\XNL$ is a closed linear subspace
of $\X(D)$. The quadratic form in (\ref{minR})  is positively defined. Then the proof follows. $\Box$

\par
{\em Proof of Theorem \ref{ThMR}.}   First, let us show that the optimal process $\w x_\rho\in \XN$ which
existence is established in Lemma \ref{lemma1R} is such that
\baa \w x_\rho=\frac{1}{1+\rho}\, P\left(\nu(x)+\w x_\rho\Ind_{\M}\right). \label{R}\eaa
\par
Let $x_\rho'=x\Ind_{\D}+\w x_\rho\Ind_{\M}$.  Then   $\w x_\rho$
is an unique solution of  the minimization problem \baaa &&\hbox{Minimize}\quad  \sum_{t\in\ZZ}\|x_\BL(t)-x_\rho'(t)\|_V^2+\rho\|x_\BL\|_{\X}^2 \quad\breakk\hbox{over}\quad x_\BL\in \XN.\label{minRPP} \eaaa
It follows that   $\w x_\rho=(1+\rho)^{-1}\ww x_\rho$, where   $\ww x_\rho$
is an unique solution of  the minimization problem \baaa &&\hbox{Minimize}\quad  \sum_{t\in\ZZ}\|x_\BL(t)-x'_\rho(t)\|_V^2 \quad\breakk\hbox{over}\quad x_\BL\in \XN.\label{minR1} \eaaa

By the property of the projection, $\ww x_\rho=P x_\rho'$.
By the definitions, it follows that
\baaa (1+\rho)\w x_\rho= \ww x_\rho=P\left(\nu(x)+x'_\rho\Ind_{\M}\right)\brea=P\left(\nu(x)+\w x_\rho\Ind_{\M}\right). \eaaa
Similarly to the proof of Theorem \ref{ThM}, we have that (\ref{R}) holds.

Further, equation (\ref{R}) is  equivalent to equation (\ref{yAaR}) which, on its turn, is  equivalent to the equation  \baaa y=A_\rho y+a_\rho (x).
\label{yR2}
\eaaa
Since the operator $(I-A_\rho)^{-1}: \X^M\to \X^M$ is continuous, this equation  has an unique solution
$y_\rho=\w x_\rho\,\Ind_{\M}=(I-A_\rho)^{-1}a_\rho(x)$ in $\X^M$, and the required estimate for $\|y_\rho\|_{\X^M}$ holds.
 This completes the proof of Theorem \ref{ThMR}. $\Box$
\par
{\em Proof of Lemma \ref{lemma4}.} Let us prove statement (i).
 We have that $A_N y=\Ind_{D_N}(P  (\Ind_{D_N}y)$.
Then statement (i) follows.  Further, it follows that $(1+\rho)^{-1}\|A_N\|<1$ for the
norm of the operator $A_N:\X^m\to\X^M$. . Then statement (i) follows.
This completes the proof of Lemma \ref{lemma4}. $\Box$
\par
 {\em Proof of Theorem \ref{Th3}}.
  Let $e_N=y_{\rho,\eta,N}-y_\rho$.  We have that
     \baa (1+\rho)e_N=A_Ne_N+ (A_N-A)y_\rho+a(x_\eta)-a(x).
   \label{AAN}  \eaa
    Hence
   \baaa
   (A_N-A)y_\rho= \Ind_{D_N}[P  (\Ind_{D_N}y_\rho )-\Ind_{\M}(P  y_\rho )]\brea=\w\zeta_{N,\rho}+\ww \zeta_{N,\rho},
   \eaaa
   where
    \baaa
   \w\zeta_{N,\rho}= \Ind_{D_N}[P  (\Ind_{D_N}y_\rho )-P  y_\rho ]=\Ind_{D_N}[P  (\Ind_{D_N}y_\rho - y_\rho) ]\\ =
   \Ind_{D_N}[P  (\Ind_{D_N}y_\rho - y_\rho) ]=-\Ind_{D_N}[P  (\Ind_{\{t:\ t>N\}}y_\rho)\eaaa
   and \baaa
   \ww\zeta_{N,\rho}= [\Ind_{D_N}-\Ind_{\M}](P  y_\rho )= -\Ind_{\{t:\ t> N\}}(P  y_\rho ).
   \eaaa
   Clearly, $\|\w\zeta_{N,\rho}\|_{\X^M}\to 0$ and  $\|\ww\zeta_{N,\rho}\|_{\X^M}\to 0$ as $N\to +\infty$. Hence  $\|(A_N-A)y_\rho\|_{\X^M}\to 0$ as $N\to +\infty$.

     By the conitinuity of the operator $a(\cdot)$, we have that  \baaa
     \|a(x)-a(x_\eta)\|_{\X^M}\le \|\eta\|_{\X(D)}.
     \eaaa
     Hence  \baaa
     &&\|e_{\rho,\eta,N}\|_{\ell_2(M,V)}\le \|(I-A_\rho)^{-1}\|\Bigl(\|(A_N-A)y_\rho \|_{\X^M}\breakk+\|a_\rho(x)-a_\rho(x_\eta)\|_{\X^M}\Bigr)
     \\ &&\le \|(I-A_\rho)^{-1}\|\Bigl(\|(A_N-A)y_\rho \|_{\X^M}+\|\eta \|_{\X(D)}\Bigr)\to 0\eaaa
  as $ N\to +\infty$ and $\|\eta\|_{\X(D)}\to 0 $. This completes the proof of Theorem \ref{Th3}. $\Box$

\def\XN{\ell_2^\BL}
\def\XNL{\ell_2^\BL(D)}
\def\BL{{\scriptscriptstyle BL}}

{\em Proof of Proposition \ref{propU}}. Let us show that Condition \ref{condD}(i) is satisfied.
It suffices to consider the case where $\{t:\ t\le 0\}\subset \D$. Furthermore, it suffices to
prove that if $x(\cdot)\in\XN $ is such that $x(t)=0$ for
$t\le 0$, then $x(t)=0$ for $t>0$. The proof of this repeats the proof of Proposition 1
\cite{D17a}.
\par  Let us show that Condition \ref{condD}(ii) is satisfied.
Consider the mapping $\zeta:\BN \to \XNL$ such that
$x(t)=(\zeta (X))(t)=(\Z^{-1} X)(t)$ for $t\in\TT$. It is a linear
continuous operator. By Condition \ref{condD}(i), it is a bijection.
\par
 Since  the mapping $\zeta:\BN \to \XNL$ is continuous, it follows that
the inverse mapping $\zeta^{-1}: \XNL\to\BN$ is also
continuous; see, e.g., Corollary in Ch.II.5 \citet{Yosida}, p. 77. Since the
set $\BN$ is a closed linear subspace of $L_2(-\pi,\pi)$, it
follows that $\XNL$ is a closed linear subspace of $\X(D)$.
Then a solution $\w x$  of problem (\ref{min})
is such that $\w x|_{D}$ is  a projection of $x|_{D}$ on $\XNL$ which is unique.
 This completes the proof of Proposition \ref{propU}.  $\Box$

{\subsection*{Acknowledgment}
This work  was supported by ARC grant of Australia DP120100928 to the author.}

\end{document}